\documentclass[preprint]{vgtc}               

\graphicspath{{figures/}{pictures/}{images/}{./}} 

\usepackage{times}                     

\usepackage{tabu}                      
\usepackage{booktabs}                  
\usepackage{lipsum}                    
\usepackage{mwe}                       
\usepackage{mathptmx}                  
\usepackage{subcaption}
\usepackage{amsmath,bm}
\usepackage{multirow}
\usepackage{enumitem}
\usepackage{xspace}
\usepackage{xcolor}

\onlineid{1017}

\vgtccategory{Research}

\preprinttext{To appear in the IEEE Transactions on Visualization and Computer Graphics (TVCG) Special Issue on IEEE ISMAR 2026.}

\title{Evaluating the Vergence–Accommodation Conflict in Gaze-Based 3D Target Selection}

\author{%
  \authororcid{Mohammad Raihanul Bashar}{0000-0002-5271-457X},
  \authororcid{Mohammadreza Amini}{0009-0009-5711-2162},
  \authororcid{Aunnoy K Mutasim}{0000-0002-5321-7292},
  \texorpdfstring{\\[-1pt]}{ }
  \authororcid{Mayra Donaji Barrera Machuca}{0000-0002-8057-7103},
  \authororcid{Wolfgang Stuerzlinger}{0000-0002-7110-5024}, and
  \authororcid{Anil Ufuk Batmaz}{0000-0001-7948-8093}
}

\authorfooter{
  \item
    Mohammad Raihanul Bashar and Mohammadreza Amini are with Concordia University.\\
  	E-mail: {mohammadraihanul.bashar | mohammadrezaamini}@mail.concordia.ca.
  \item
    Aunnoy K Mutasim and Mayra Donaji Barrera Machuca are with the University of Calgary.\\
  	E-mail: {aunnoy.mutasim | mbarrera}@ucalgary.ca.
  \item
    Wolfgang Stuerzlinger is with Simon Fraser University.\\
  	E-mail: w.s@sfu.ca.
  \item
  	Anil Ufuk Batmaz is with Concordia University.\\
  	E-mail: ufuk.batmaz@concordia.ca.
}

\teaser{
  \centering
  \includegraphics[
    width=\linewidth,
    alt={Two-part illustration of the experimental setup. Left: targets are positioned at multiple depths spaced in diopters while maintaining a constant perceived size. The angular target distance alpha is shown in orange, the angular target width omega is shown in blue, and dagger symbols indicate the focal planes of the head-mounted displays. Right: a participant selects virtual targets using either a handheld controller or gaze, with the active cursor highlighted.}
    ]{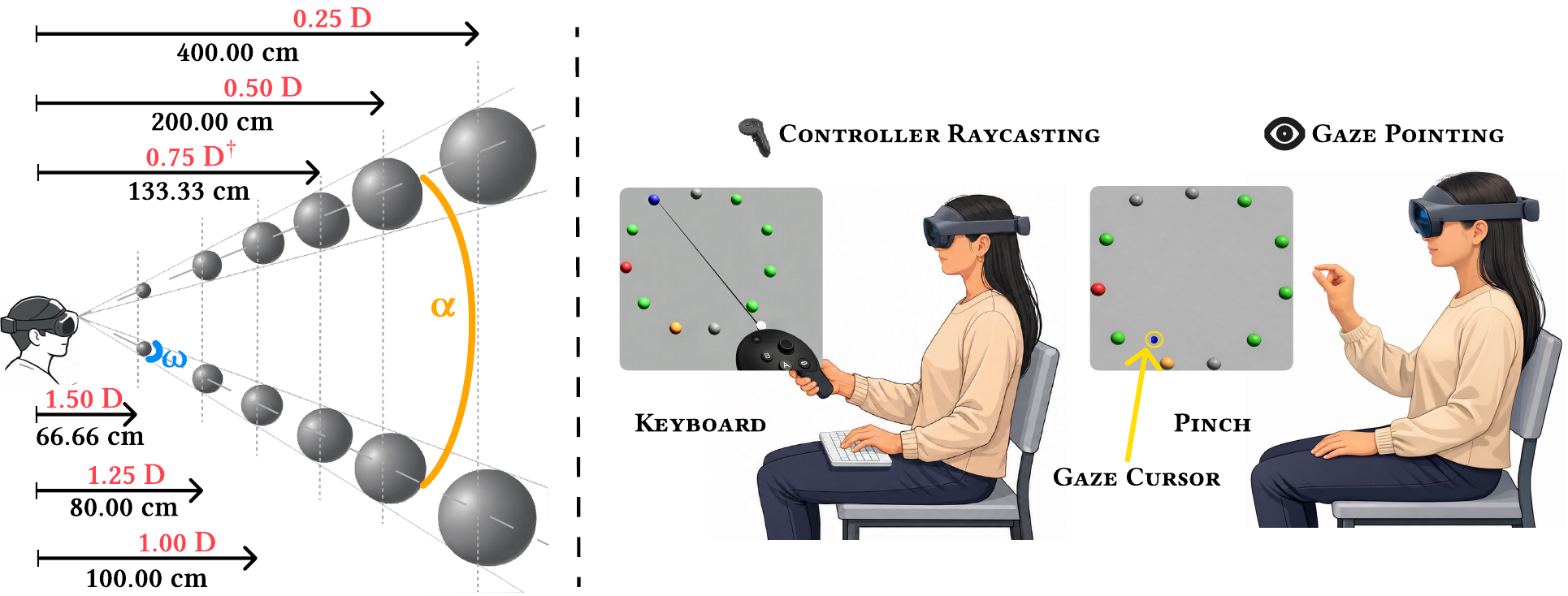}
  \caption{Experimental task and depth configuration. (Left) Targets were placed at multiple depths spaced in diopters while maintaining constant perceived size. $\alpha$ (orange) denotes angular target distance and $\omega$ (blue) the angular target width~\cite{kopper2010human}. A dagger (†) marks the HMDs' focal plane. (Right) Participants selected targets using the controller or gaze (cursor highlighted).}
  \label{fig:teaser}
}

\newcommand{\gaze}{\textsc{Gaze}\xspace}
\newcommand{\controller}{\textsc{Controller}\xspace}

\abstract{State-of-the-art head-mounted displays (HMDs) enable gaze-based selection in virtual environments. Yet, these HMDs suffer from the vergence–accommodation conflict (VAC), which is known to affect interaction performance. The VAC might influence gaze-based selection performance because it directly affects eye-movement behavior. Thus, in this paper, we investigate how the VAC influences \gaze-based 3D target selection across varying depth conditions. Our results show that as the (visual) depth increases, user performance significantly decreases with gaze-based selection. Moreover, a previously suggested ``\textit{Variation in Diopter}'' Fitts’ law model captured this performance change better relative to a linear model. These findings provide evidence that gaze-based pointing is negatively affected by the VAC and highlight the importance of accounting for depth-dependent factors when designing gaze-based interaction in 3D environments.} 


\begin{document}




\firstsection{Introduction}

\maketitle

Selection is a fundamental interaction task in extended reality (XR) systems~\cite{laviola20173d}.~Many XR applications rely on the ability to select objects in three-dimensional (3D) space, for activating menu elements, selecting virtual objects, or teleporting to or interacting with distant content.~In these scenarios, users must first visually identify a target, estimate its spatial location, and perform the corresponding selection action.~Successful target selection, therefore, depends on the coordination of perceptual processes, such as depth perception and spatial judgment, with motor execution. Thus, users must continuously interpret depth cues and spatial relationships when performing interaction tasks. Prior research has shown that 3D selection performance is strongly influenced by how depth is represented by the display and interpreted by the user \cite{kramida2015resolving, vienne2020depth}. In 3D interaction, these limitations can affect users’ ability to judge distances and accurately select targets~\cite{bashar2025effect, batmaz2022effect}.

Human depth perception relies on a combination of monocular and binocular cues~\cite{armbruster2008depth}. Contemporary stereoscopic head-mounted displays (HMDs) reproduce several cues used for depth perception, but they typically present imagery on a fixed focal plane while rendering virtual objects at varying depths. This design leads to one of the most widely studied perceptual limitations of immersive displays: the vergence–accommodation conflict (VAC)~\cite{adams2022depth, kramida2015resolving, pfeil2021distance, zhou2021vergence}.~In natural viewing, vergence (the rotation of the eyes) and accommodation (the focusing of the eye lens) are tightly coupled, allowing the eyes to converge and focus at the same physical distance.~In contrast as stereoscopic XR HMDs render images on a fixed plane while simulating objects at varying virtual depths using binocular disparity~\cite{zhou2021vergence}. As a result, users must accommodate to the display plane while converging to different virtual depths, creating a mismatch between vergence and accommodation cues~\cite{banks2013insight, shibata2011zone, zhou2021vergence}. Prior research has shown that the VAC can impair depth perception, increase visual discomfort and fatigue, and negatively affect performance in spatial tasks~\cite{hoffman2008vergence, lin2022effects, mcanally2024vergence}. In 3D interaction, these perceptual distortions can directly affect users’ ability to judge distances and accurately acquire targets, ultimately affecting selection performance~\cite{bashar2025effect, batmaz2022effect}.

However, prior work examining the effects of the VAC on interaction has primarily focused on hand-based selection techniques, such as controller-based raycasting or virtual hand interaction \cite{barrera2019effect, batmaz2022effect, batmaz2023re}.~While these studies provide valuable insights into how VAC-related perceptual limitations influence manual pointing performance, their results cannot be directly translated to gaze-based pointing.~Unlike manual techniques, which rely on hand motion, gaze-based pointing depends directly on oculomotor behavior, including vergence mechanisms involved in binocular fixation.

In gaze-based pointing, users typically fixate on a target before initiating a confirmation action \cite{mutasim2020gaze, mutasim2021pinch, lin2022effects}.~More broadly, gaze plays a fundamental role in guiding spatial interactions, even when selection is ultimately performed with the hand or a controller \cite{sidenmark2019gaze}.~With the increasing availability of eye tracking in modern HMDs, gaze has also emerged as an explicit pointing modality for target selection \cite{luro2019comparative, mutasim2021pinch, pfeuffer2024design, pfeuffer2017gaze+}.~As, gaze pointing depends directly on oculomotor behavior involved in binocular fixation, VAC-related changes in vergence demand may directly affect gaze-driven selection performance.~This motivates the following research question: \textit{How does the vergence–accommodation conflict affect gaze-based 3D target selection performance?}

To address this question, we conducted a user study in an immersive virtual environment where participants performed a multidirectional 3D target selection task using gaze-based pointing, with controller raycasting included as a reference condition.~Targets were presented at six depth levels, expressed in diopters and separated by 0.25 increments, to examine gaze-based selection performance under varying vergence–accommodation demands.~Our findings show that gaze performance varies significantly across depth, with reduced performance as targets move farther from the display focal region. Furthermore, a depth-aware Fitts’ law model that accounts for diopter variation captures these performance changes more effectively than the standard formulation. Based on these findings, we discuss design implications and provide recommendations for depth-aware gaze interaction in XR.

\section{Related Work}

\subsection{Vergence-Accommodation Conflict and Depth Perception}

The VAC is a well-known limitation of stereoscopic HMDs, caused by a mismatch between the vergence-driven depth cues and the fixed focal plane of the display~\cite{kramida2015resolving, zhou2021vergence}. Under natural viewing conditions, vergence and accommodation are tightly coupled, with the eyes converging and focusing at the same physical distance. In stereoscopic displays, however, users must accommodate to the display plane while converging to virtual depths defined by binocular disparity. As a result, this mismatch can impair depth perception, increase visual discomfort, and influence oculomotor behavior during interaction~\cite{chen2015visual, hoffman2008vergence, read2016viewing, shibata2011zone}.

Prior research has shown that the VAC can reduce depth discrimination accuracy~\cite{hoffman2008vergence}, distort perceived depth~\cite{shibata2011zone}, and increase visual fatigue during prolonged viewing~\cite{chen2015visual, read2016viewing}.~It can also disrupt oculomotor behavior, including increased vergence latency, reduced convergence accuracy, and overshoot during fixation shifts when targets deviate from the display focal plane~\cite{daniel2019induced, sumarokova2024individual, vienne2014effect}.~Together, these effects may destabilize eye movements when fixating objects located far from the display focal region~\cite{diez2019effect}.

Vision science identifies a \textit{zone of comfort} for stereoscopic viewing, typically spanning $\pm 0.5$ diopters around the focal plane~\cite{shibata2011zone}.~Within this zone, the visual system can maintain stable binocular vision with minimal effort; outside it, vergence-accommodation mismatch can reduce image quality, increase visual fatigue, and decrease viewing comfort~\cite{kramida2015resolving}.~These limitations have motivated rendering approaches that provide additional focus cues in immersive displays. Prior work has shown that defocus blur can influence perceived distance and scale~\cite{t2010using}, gaze-contingent depth-of-field rendering can enhance perceived depth and realism~\cite{mauderer2014depth}, and ChromaBlur leverages longitudinal chromatic aberration for depth-dependent blur cues that drive accommodation and improve perceived depth~\cite{cholewiak2017chromablur}. Recent work has also explored inverse-blurring techniques to compensate for the VAC~\cite{hussain2023improving} and compared correct versus simulated focus cues in stereoscopic scenes~\cite{march2024impact}.

In parallel, several display hardware technologies have been proposed to mitigate the VAC, including multifocal displays~\cite{batmaz2022effect, zhan2020multifocal}, varifocal systems~\cite{dunn2017wide, hu2026varifocaldisplaysreduceimpact}, light-field displays~\cite{maeda2024wide, panzirsch2022light}, and holographic displays~\cite{park2022holographic, rathinavel2019varifocal}. These approaches aim to restore the natural coupling between vergence and accommodation by providing appropriate accommodation cues~\cite{chang2020toward, zhou2021vergence}. However, most commercial XR systems still use conventional fixed-focus stereoscopic displays, meaning users remain subject to VAC-related perceptual limitations.

\subsection{Effects of VAC on 3D Selection}

Beyond perceptual effects, research shows that the VAC also influences performance in interactive tasks within immersive environments. In particular, VAC-related depth perception errors have been shown to impair performance in 3D pointing and target selection tasks~\cite{bashar2025effect, batmaz2022effect, fernandes2025looking}. Inaccuracies in depth perception can directly affect motor performance during target acquisition, because many interaction techniques require users to estimate object distances and align a cursor, ray, or virtual hand with targets in 3D space.

Early work by Barrera and Stuerzlinger~\cite{barrera2019effect} showed that pointing movements involving depth changes are significantly slower than lateral movements in stereoscopic displays, suggesting that depth-perception limitations affect the planning and execution of pointing actions~\cite{barrera2018stereo}. Subsequent studies by Batmaz et al.~\cite{batmaz2023re, batmaz2023virtualhandvac} further reported that movements along the depth axis exhibit lower throughput and require more corrective sub-movements than lateral movements, indicating increased interaction difficulty when targets are distributed across depth.

Similar effects have been reported for ray-based pointing techniques, where variations increase movement time, reduce throughput, and increase pointing variability~\cite{bashar2025effect, batmaz2023virtualhandvac, batmaz2023re}. These performance changes are often attributed to increased uncertainty in depth estimation, which leads users to perform additional corrective movements during the final stages of pointing~\cite{wang2024geometry}.

Importantly, most studies investigating VAC-related interaction effects have focused on manual pointing techniques, such as controller-based raycasting or virtual hand interaction~\cite{bashar2025effect, batmaz2023re}. In these cases, VAC primarily affects manual motor behavior, including longer ballistic phases, more corrective movements, and reduced movement efficiency~\cite{barrera2019effect, batmaz2022effect, itoh2021towards}. In contrast, relatively little is known about how VAC influences interaction techniques that rely directly on eye movements, such as gaze-based pointing.

\subsection{Gaze-Based Pointing and Performance Modeling in 3D Interfaces}

Gaze-based pointing has long been explored in 2D interfaces and has recently gained attention in 3D and immersive environments. Compared to manual pointing, gaze input can reduce physical effort, support faster target acquisition, and directly align visual attention with selection~\cite{stellmach2012look, zhai1999manual}. As users typically look at targets before interacting with them, gaze can also serve as a natural indicator of intent in spatial interfaces~\cite{jacob1990you}. In gaze-based interaction, eye fixation is often used to indicate a target, while selection is confirmed through dwell or a explicit action such as a button press or pinch gesture~\cite{miniotas2000application, mutasim2021pinch}.~Prior work has examined gaze selection speed and accuracy~\cite{miniotas2000application, stellmach2012look}, the \textit{Midas touch} problem~\cite{jacob1990you}, and the role of dwell-time thresholds and confirmation mechanisms~\cite{miniotas2000application, mutasim2025there, mutasim2022saccades, mutasim2021pinch}, showing that gaze can be effective when paired with explicit confirmation to reduce unintended selections~\cite{mutasim2022saccades, pfeuffer2024design, pfeuffer2017gaze+}.

Despite these advantages, gaze pointing presents challenges.~Eye movements are inherently fast~\cite{bahill1975main} but can be less spatially precise than manual motor actions, leading to increased pointing variability and reduced accuracy in some interaction scenarios~\cite{boukhelifa2019exploratory, hansen2018fitts, mott2017improving}. Prior work has therefore explored techniques to stabilize gaze input, filter eye-movement noise~\cite{feit2017}, and combine gaze with other input modalities to improve selection performance~\cite{bashar2026eyesonmany, pfeuffer2024design}.

From a physiological perspective, gaze behavior is closely tied to oculomotor mechanisms such as vergence and accommodation, which maintain binocular fixation across depths.~Vision science has shown that the VAC can alter vergence dynamics and reduce fixation stability during stereoscopic viewing~\cite{banks2013insight, batmaz2022effect, hoffman2008vergence}. Specifically, the VAC has been linked to increased vergence latency, fixation overshoot during gaze shifts, and unstable eye movements for targets far from the display focal plane~\cite{daniel2019induced, sumarokova2024individual}. Despite this, gaze tracking has primarily been studied as a measurement of visual attention rather than as an active input modality affected by the VAC.

To quantitatively evaluate pointing performance, HCI research commonly relies on Fitts’ law, which models the relationship between movement time and task difficulty~\cite{mackenzie1992fitts}. The Shannon formulation is defined as:
\begin{equation}
MT = a + b \cdot \mathit{ID} \quad
with \quad\mathit{ID} = \log_2\left(\frac{A}{W} + 1\right)
\end{equation}

\noindent where $MT$ is the movement time, $A$ movement amplitude, $W$ target width, and $\mathit{ID}$ is the index of difficulty. The constants $a$ and $b$ are empirically determined through regression and represent baseline movement time and sensitivity to task difficulty, respectively.~As Fitts’ law characterizes the speed--accuracy trade-off in pointing tasks, it has been widely used to evaluate interaction modalities such as mouse and touch input or gaze-based pointing~\cite{hansen2018fitts, medeiros2023benefits, miniotas2000application}.

In summary, prior work shows that the VAC can degrade depth perception and affect selection performance in manual 3D pointing tasks, while gaze-based interaction has emerged as an important input modality in immersive systems. However, little is known about how VAC influences gaze-driven target selection, despite gaze relying directly on oculomotor mechanisms such as vergence and accommodation. This work, therefore, investigates how the VAC affects gaze-based 3D target selection across varying depths.

\section{Motivation}

Gaze is increasingly used as an input modality in modern XR systems, yet prior studies on VAC-related interaction effects have largely focused on manual techniques such as controller-based raycasting~\cite{barrera2018stereo, Batmaz2022VRST}. These findings mainly characterize how depth uncertainty affects visually guided hand or controller movements. In contrast, gaze-based pointing depends directly on oculomotor mechanisms involved in binocular fixation~\cite{hoffman2008vergence, monier2025ocular, rocchi2025comparison}. As the VAC can affect vergence responses and fixation stability, manual-pointing results may not generalize to gaze-driven interaction. This is important because XR interfaces often place interactive elements at different depths, where VAC-related effects may make gaze-selectable targets slower or less accurate to select. Thus, we examined the following \textbf{research question:} \textit{How does target depth influence gaze-based selection performance under the vergence-accommodation conflict in XR environments?}

\section{User Study}

This study investigates how VAC-related depth variations affect gaze-based 3D target selection performance.~This choice is also consistent with prior 3D selection literature, which identifies depth as a recurring task-environment factor used to evaluate how spatial layout affects selection performance~\cite{bashar2024virtual}.~Specifically, participants selected targets at different visual depths relative to the display’s focal plane using an ISO 9241:411 multidirectional selection task~\cite{ISO2015}.

Unlike prior work~\cite{barrera2019effect, batmaz2023virtualhandvac, Cha2013c, Murata2001c} that expressed depth in linear distance, we represent depth changes in diopters, a measure of optical focus that corresponds to the eye’s accommodative depth and is commonly used to describe depth under the VAC. Recent work shows that diopter-based modeling better captures the visual demands of stereoscopic displays~\cite{bashar2025effect}. While prior studies focused on controller-based pointing, we apply this representation to gaze-based selection.

\subsection{Methodology}

\subsubsection{Participants}

We conducted an a priori power analysis using G*Power~\cite{faul2007g}, which indicated that 12 participants would provide 80\% statistical power to detect a large effect ($\eta^2 = 0.14$) at $\alpha = .05$ using a repeated-measures (RM) ANOVA. To account for potential variability, we recruited 24 participants (13 female, 11 male)~\cite{bergstrom2021evaluate}, aged 20 -- 28 years ($M = 23.16$, $SD = 1.75$). All were right-handed; 21 reported right-eye dominance and 3 left-eye dominance. Nineteen had normal vision, and five had corrected-to-normal vision; none reported color vision deficiencies or binocular vision impairments.~Six participants had never used VR, five had used it 1 -- 5 times, and thirteen reported more than 5 VR experiences. Only seven had prior gaze-based interaction experience. The study protocol was approved by the institutional ethics board.

\subsubsection{Apparatus}

The study was conducted on a PC with an Intel® Core™ i9-13900KF processor (3.0\, GHz), 32\, GB RAM, and an NVIDIA GeForce RTX\,4070 GPU. We used a Meta Quest Pro HMD with a focal distance of 1.33 m \cite{johncarmack}, per-eye resolution of 2064\(\times\)2208 pixels, and a horizontal and vertical FOV of 106$^{\circ}$ and 96$^{\circ}$, respectively, and a $90~\mathit{Hz}$ refresh rate. The HMD provides integrated binocular video-based eye tracking sampled at $72~ \mathit{Hz}$~\cite{rocchi2025comparison}; prior evaluations report spatial accuracy of approximately 1.2$^{\circ}$ -- 1.8$^{\circ}$ and precision of approximately 0.7$^{\circ}$ -- 0.9$^{\circ}$~\cite{aziz2024evaluation, wei2023preliminary}. We used the SDK-provided combined gaze ray for gaze input and placed the gaze cursor at the first intersection between this ray and the virtual scene geometry. The virtual environment was developed in Unity 2021.3.24 using the Oculus Unity Integration SDK (v57.0.1-deprecated) and displayed in the HMD via Meta Quest Link.

\begin{figure*}
    \centering
    \includegraphics[
        width=1\linewidth,
        alt={Flowchart of the user study procedure. Participants first provided informed consent, then completed a training phase, performed the experimental tasks, and finally completed post-study questionnaires.}
    ]{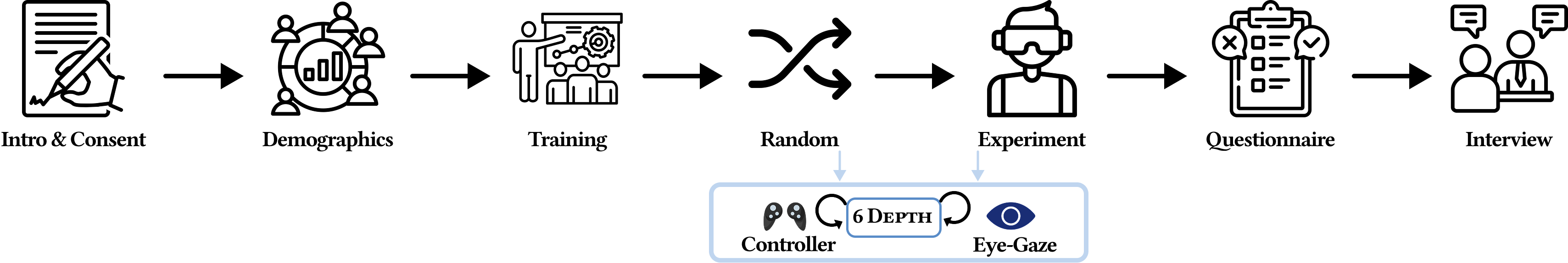}
    \caption{Overview of the user study's experimental procedure, including consent, training, task execution, and post-study questionnaires.}
    \label{fig:exp}
\end{figure*}

\subsubsection{User Study Procedure}

Before the experiment, participants completed a consent form and demographic questionnaire. An experimenter then explained the procedure and assisted participants with donning the HMD and completing eye-tracking calibration. Before the main trials, participants completed a short training or familiarization session to become accustomed to the task, visual feedback, pointing modalities, and confirmation methods, reducing initial unfamiliarity with gaze-based 3D selection~\cite{mutasim2025influence}. Participants performed the task using both gaze-based pointing and controller-based raycasting, with condition order counterbalanced using a Latin Square design. In the \controller condition, participants held the controller in their dominant hand with their arm close to their body; a virtual ray extended from the controller, and a spherical cursor appeared at its intersection with the scene geometry. The cursor subtended a constant visual angle of approximately 0.21$^{\circ}$ across depth conditions, corresponding to a 0.5\,cm diameter at the HMD focal plane (133.33\,cm), and was scaled with distance to avoid apparent-size differences across depths. In the \gaze condition, the same cursor followed the participant’s gaze direction. The confirmation method differed by modality to minimize motor interference: \controller selections were confirmed by pressing the space bar with their non-dominant hand, mitigating the ``\textit{Heisenberg effect}''~\cite{bowman2001using}, while \gaze selections were confirmed using a dominant-hand pinch gesture. Prior work shows that pinch and click confirmations yield comparable performance and generally outperform dwell-based selection~\cite{mutasim2021pinch}.

Participants performed a multidirectional selection task based on the ISO~9241-411~\cite{ISO2015}. Targets were arranged as 11 gray spheres in a circular layout (see~\Cref{fig:teaser}); one sphere was highlighted in orange at a time, and participants selected it as quickly and accurately as possible.~The first target was chosen randomly, and subsequent targets alternated across the circle in either clockwise or counterclockwise order.~When the cursor approached within 0.5\,cm of a sphere, the sphere turned blue to provide feedback~\cite{teather2014visual}; successful selections turned it green, while errors turned it red and triggered an auditory cue~\cite{batmaz2021pitch}. The task was performed seated in a static virtual room with a uniform grid background for spatial reference. After each set of 11 selections, the target layout was realigned relative to the participant's head position to keep the targets in front of the user.~Breaks were offered between rounds, although none of the participants chose to take one.

To study depth effects, targets were presented at different distances relative to the display focal plane. Depth was expressed in diopters ($D = 100/\text{distance (cm)}$), ranging from $1.50\,D$ to $0.25\,D$ in increments of $0.25\,D$ (\Cref{fig:teaser}). Targets were placed in front of, at, and behind the Quest Pro's 133.33\, cm focal distance, i.e., $0.75\,D$~\cite{johncarmack}.

Following prior work~\cite{barrera2018stereo, bashar2026effects, bashar2025effect, medeiros2016perceiving}, the target size was adjusted with distance to maintain a constant visual angle across depth conditions. After completing the experiment, participants completed the System Usability Scale (SUS)~\cite{Brooke1996SUS} and NASA-TLX~\cite{NASA-TLX} questionnaires and provided qualitative feedback on their experience and preferred interaction modality.~The experiment lasted approximately 35 -- 40 minutes per participant (see~\Cref{fig:exp}).

\begin{table*}
\centering
\caption{RM ANOVA results showing the effects of Depth, ID, and their interaction. Significant results are shown in bold.}
\label{tab:RM-ANOVA}
\resizebox{0.7\textwidth}{!}{%
\begin{tabular}{cccc}
\hline
                   & Depth                                                                                                          & ID                                                                                                     & Depth * ID                                                                                             \\ \hline
Movement Time      & \begin{tabular}[c]{@{}c@{}}$\bm{F(3.46, 79.585) = 75.85,}$ \\ $\bm{p < .001, \eta_p^2 = .767}$\end{tabular}    & \begin{tabular}[c]{@{}c@{}}$\bm{F(8, 184) = 51.177,}$ \\ $\bm{p < .001, \eta_p^2 = .690}$\end{tabular} & \begin{tabular}[c]{@{}c@{}}$\bm{F(40, 920) = 7.864,}$ \\ $\bm{p < .001, \eta_p^2 = .255}$\end{tabular} \\ \hline
Error Rate         & \begin{tabular}[c]{@{}c@{}}$\bm{F(5, 115) = 2.462,}$ \\ $\bm{p < .037,\eta_p^2 = .097}$\end{tabular}           & \begin{tabular}[c]{@{}c@{}}$\bm{F(8, 184) = 2.410,}$ \\ $\bm{p < .017, \eta_p^2 = .095}$\end{tabular}  & \begin{tabular}[c]{@{}c@{}}$\bm{F(40, 920) = 2.095,}$ \\ $\bm{p < .032, \eta_p^2 = .086}$\end{tabular} \\ \hline
Angular Throughput & \begin{tabular}[c]{@{}c@{}}$\bm{F(5, 115) = 18.653,}$ \\ $\bm{p < .001, \eta_p^2 = .448}$\end{tabular}          & \begin{tabular}[c]{@{}c@{}}$\bm{F(8, 184) = 4.769,}$ \\ $\bm{p < .001,\eta_p^2 = .172}$\end{tabular}   & \begin{tabular}[c]{@{}c@{}}$\bm{F(40, 920) = 2.560,}$ \\ $\bm{p < .001, \eta_p^2 = .13}$\end{tabular}  \\ \hline
$\mathit{SD}_x$    & \begin{tabular}[c]{@{}c@{}}$\bm{F(1.536, 35.321) = 122.037,}$ \\ $\bm{p < .001, \eta_p^2 = .841}$\end{tabular} & \begin{tabular}[c]{@{}c@{}}$\bm{F(8, 184) = 28.682,}$ \\ $\bm{p < .001, \eta_p^2 = .555}$\end{tabular} & \begin{tabular}[c]{@{}c@{}}${F(40, 920) = .717,}$ \\ ${p = .906, \eta_p^2 = .030}$\end{tabular}        \\ \hline
\end{tabular}%
}
\end{table*}

\subsubsection{Experimental Design}

We used a within-subjects design with two independent variables: \textit{Pointing Modality} (\controller and \gaze) and \textit{Depth}. We used the same depth levels as Bashar et al.~\cite{bashar2025effect} to ensure methodological consistency and facilitate comparison with prior work.

Consistent with prior depth-aware pointing studies~\cite{bashar2025effect, batmaz2022effect}, we collected the following dependent variables: \textbf{Movement Time}, the time (in seconds) from the moment a target appeared (highlighted) until the participant confirmed the selection. \textbf{Error Rate}, the number of incorrect selections. \textbf{Angular Throughput}, a combined measure of speed and accuracy derived from Fitts’ law, calculated using angular movement distance ($\alpha$) and angular target width ($\omega$). Task difficulty was expressed using the angular index of difficulty ($\mathit{ID_A}$), defined as:
\begin{equation}
\mathit{ID_A} = \log_2\left(\frac{\alpha}{\omega} + 1\right)
\end{equation}

\noindent where $\alpha$ denotes the angular movement amplitude and $\omega$ denotes angular target width. Following Kopper et al.~\cite{kopper2010human}, these angular measures maintain consistent task difficulty across viewing distances. This ensured that performance differences were attributable to depth variation rather than perceived target size. Angular throughput ($\mathit{TP}$), was computed as:
\begin{equation}
\mathit{TP} = \frac{\mathit{ID_e}}{MT}
\end{equation}

\noindent where $\mathit{ID_e}$ is the effective index of difficulty and $MT$ the mean movement time. Representing spatial selection accuracy, $\bm{SD_x}$ was the standard deviation of selection points relative to the target center along the primary task axis.

To vary task difficulty, we used nine levels of $\mathit{ID}_A$, derived from all combinations of three \textbf{angular target sizes} ($3_\mathit{ATS}$) and three \textbf{angular target distances} ($3_\mathit{ATD}$). We used the same angular target sizes and distances as those reported in prior research~\cite{bashar2025effect}; the complete set of values is provided in the supplementary material. Each participant completed $6_\mathit{Depth} \times 9_\mathit{ID_A} \times 2_\mathit{PointingModality} \times 11_\mathit{repetition}$, resulting in $1,188$ trials per participant and $28,512$ recorded trials overall.

\section{Results}

We preprocessed and visualized the data using JMP and conducted statistical analyses using two-way RM ANOVA in SPSS 29.0. Normality was assessed using skewness (S) and kurtosis (K), with values within $\pm1$ considered normally distributed \cite{Hair2014multivariate, Mallery2003spss}. When normality was violated, we first applied a log transformation; if violations remained, we applied the Aligned Rank Transform (ART) \cite{leys2010nonparametric, wobbrock2011aligned} to the original data and conducted the ANOVA using ARTool. We confirmed that the data met ART requirements and that non-aligned effects were removed as required by the procedure. Post-hoc comparisons used Bonferroni correction or Sidak correction for ART analyses, and Greenhouse-Geisser correction was applied when $\varepsilon < 0.75$. In all figures, error bars denote 95\% confidence intervals, and significance is annotated as $*$ ($p < .05$), $**$ ($p < .01$), and $***$ ($p < .001$). For depth-related figures, panel (a) reports results collapsed across \gaze~and \controller, panel (b) shows the two modalities separately, and panel (c) compares modalities collapsed across depth. The dagger symbol ($\dagger$) denotes the HMD focal plane at $0.75\,D$. Detailed results are presented in \Cref{tab:RM-ANOVA}; we focus the reporting on effects relevant to the VAC.

\subsection{Movement Time (MT)}

\begin{figure*}[hbt!]
    \centering
    \begin{subfigure}[t]{0.3\textwidth}
        \includegraphics[
            height=0.75\linewidth,
            alt={Plot of movement time across the tested depth conditions, measured in diopters and aggregated across interaction modalities.}
        ]{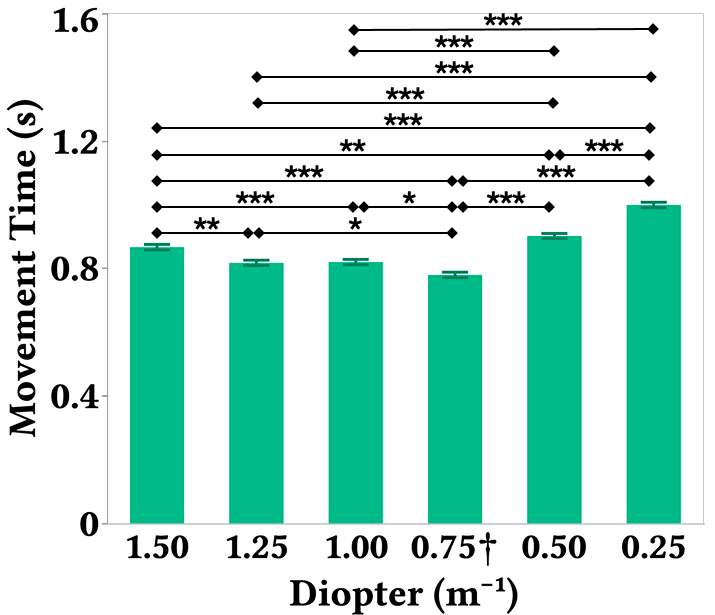}
        \caption{}
        \label{fig:MT_D}
    \end{subfigure}
    \begin{subfigure}[t]{0.3\textwidth}
        \includegraphics[
            height=0.75\linewidth,
            alt={Plot comparing movement time for gaze and controller interaction across the tested depth conditions, measured in diopters.}
        ]{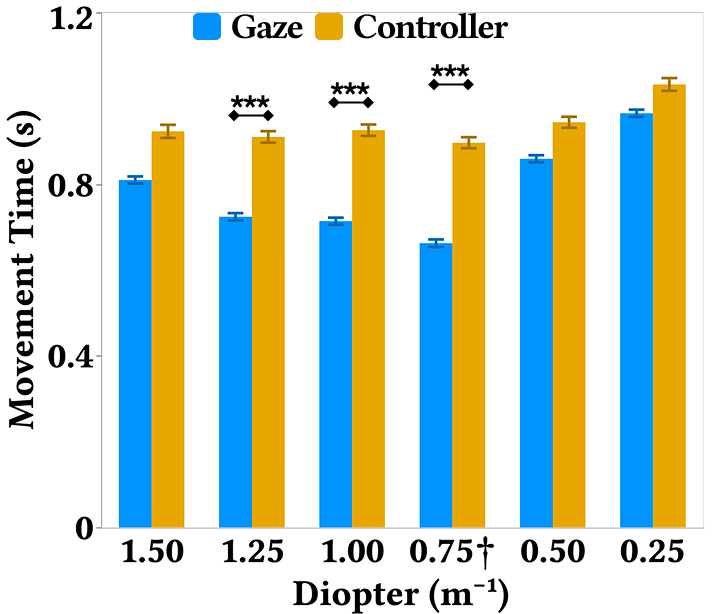}
        \caption{}
        \label{fig:MT_DC}
    \end{subfigure}
    \begin{subfigure}[t]{0.3\textwidth}
        \includegraphics[
            height=0.75\linewidth,
            alt={Plot of movement time grouped by interaction modality, comparing gaze and controller performance across depth conditions.}
            ]{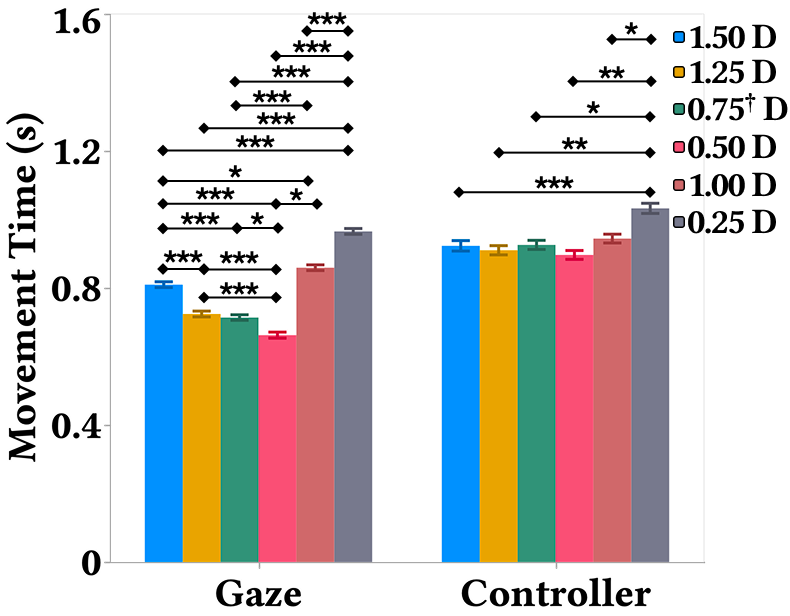}
        \caption{}
        \label{fig:MT_CD}
    \end{subfigure}
    \caption{Movement time (MT) across depth conditions (diopters) and interaction modalities. (a) MT across depths, (b) MT for \gaze and \controller across depths, (c) MT grouped by modality.}

    \label{fig:MT}
\end{figure*}

MT was normally distributed after log transformation $(S = 0.52, K = 0.94)$. A RM ANOVA revealed a significant main effect of \textbf{Depth}, $F(3.460, 79.585) = 75.850, p < .001, \eta_p^2 = .767$, (see \Cref{tab:RM-ANOVA} and \Cref{fig:MT}). Movement time increased at larger viewing distances, corresponding to lower diopter values. We also observed a significant main effect of \textbf{Pointing Modality}, $F(1, 23) = 10.077, p < .001, \eta_p^2 = .305$, with \controller pointing producing longer movement times than \gaze pointing across depth levels. Furthermore, \textbf{ID} significantly affected movement time, with higher ID values producing longer movement times. Post-hoc comparisons with Bonferroni correction revealed significant differences between several depth conditions (see \Cref{fig:MT}). For gaze-based pointing, movement time decreased as targets approached the focal region (around $0.75\,D$), but increased again for targets farther from the focal plane. In contrast, controller-based pointing showed comparatively smaller depth-dependent variation.

\begin{figure*}[hbt!]
    \centering
    \begin{subfigure}[t]{0.3\textwidth}
        \includegraphics[
            height=0.75\linewidth,
            alt={Plot of angular throughput across the tested depth conditions, measured in diopters and aggregated across interaction modalities.}
            ]{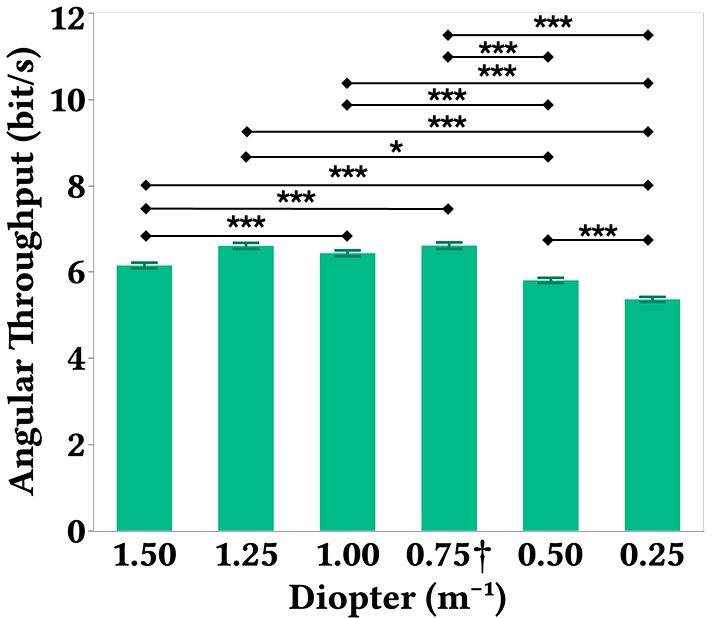}
        \caption{}
        \label{fig:THP_D}
    \end{subfigure}
    \begin{subfigure}[t]{0.3\textwidth}
        \includegraphics[
            height=0.75\linewidth,
            alt={Plot comparing angular throughput for gaze and controller interaction across the tested depth conditions, measured in diopters.}
            ]{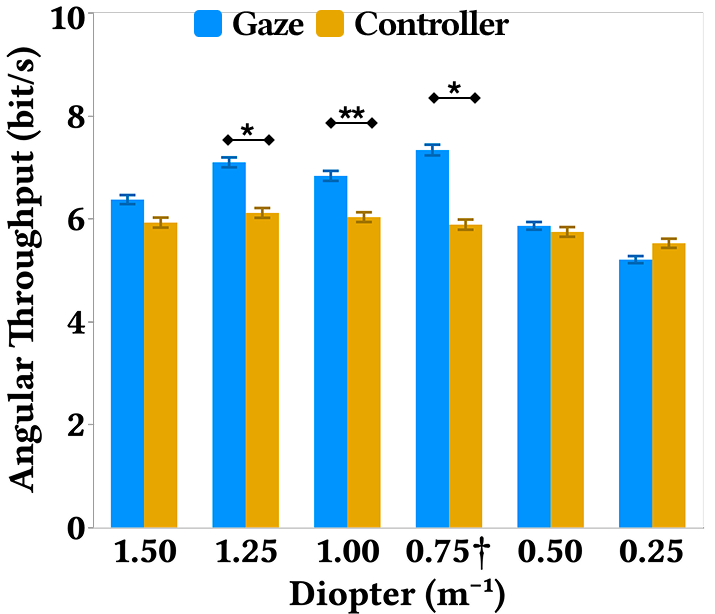}
        \caption{}
        \label{fig:THP_DC}
    \end{subfigure}
    \begin{subfigure}[t]{0.3\textwidth}
        \includegraphics[
            height=0.75\linewidth,
            alt={Plot of angular throughput grouped by interaction modality, comparing gaze and controller performance across depth conditions.}
        ]{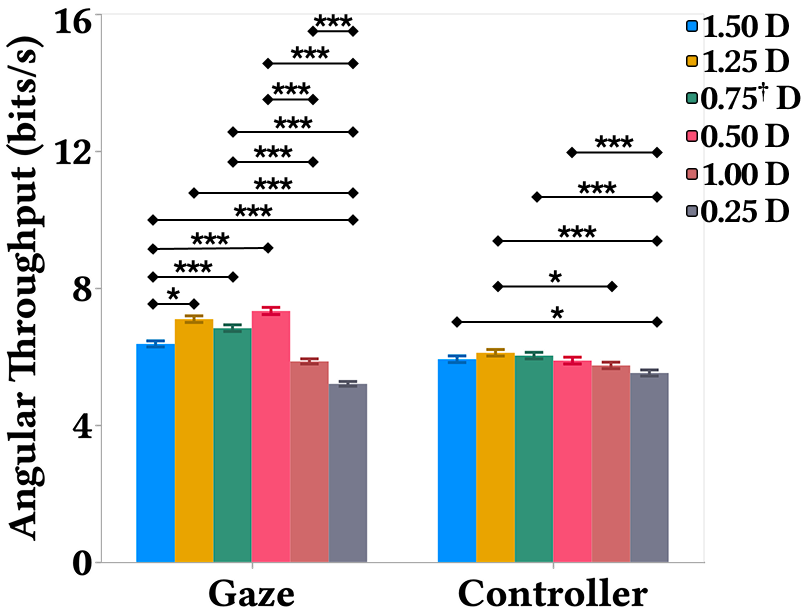}
        \caption{}
        \label{fig:THP_CD}
    \end{subfigure}
    \caption{Angular throughput across depth conditions (diopters) and interaction modalities. (a) Throughput across depths, (b) throughput for \gaze and \controller across depths, (c) throughput grouped by modality.}
    \label{fig:THP}
\end{figure*}

\begin{figure*}[hbt!]
    \centering
    \begin{subfigure}[t]{0.3\textwidth}
        \includegraphics[ height=0.75\linewidth, alt={Plot of pointing variability along the task axis, measured as SD x, across the tested depth conditions in diopters and aggregated across interaction modalities.}]{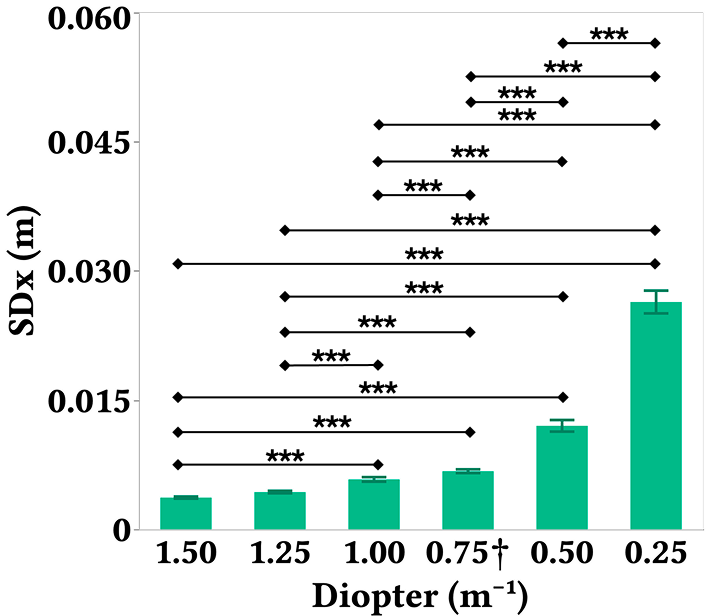}
        \caption{}
        \label{fig:SD_D}
    \end{subfigure}
    \begin{subfigure}[t]{0.3\textwidth}
        \includegraphics[ height=0.75\linewidth, alt={Plot comparing pointing variability along the task axis for gaze and controller interaction across the tested depth conditions in diopters.}]{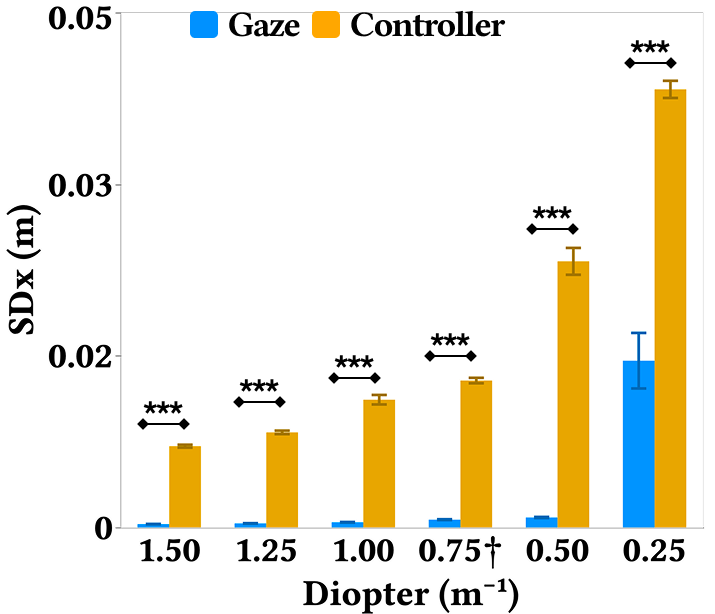}
        \caption{}
        \label{fig:SD_DC}
    \end{subfigure}
    \begin{subfigure}[t]{0.3\textwidth}
        \includegraphics[ height=0.75\linewidth, alt={Plot of pointing variability along the task axis grouped by interaction modality, comparing gaze and controller performance across depth conditions.} ]{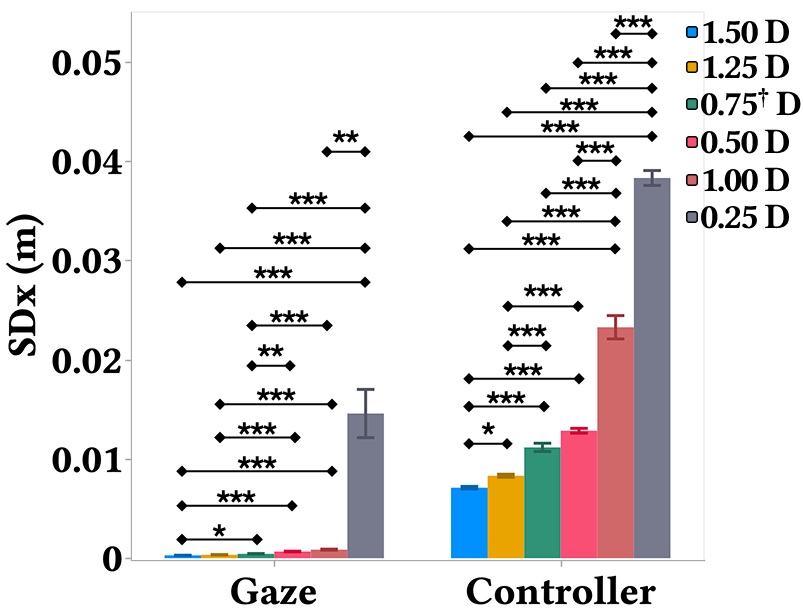}
        \caption{}
        \label{fig:SD_CD}
    \end{subfigure}
    \caption{Pointing variability along the task axis ($SD_x$) across depth conditions (diopters) and interaction modalities. (a) $SD_x$ across depths, (b) $SD_x$ for \gaze and \controller across depths, (c) $SD_x$ grouped by modality.}
    \label{fig:SD}
\end{figure*}

\begin{figure*}[t]
    \centering
    \subcaptionbox{\label{fig:ER}}{%
        \includegraphics[ height=0.18\textheight, keepaspectratio, alt={Plot of error rate across the tested depth conditions in diopters, comparing gaze and controller interaction.}]{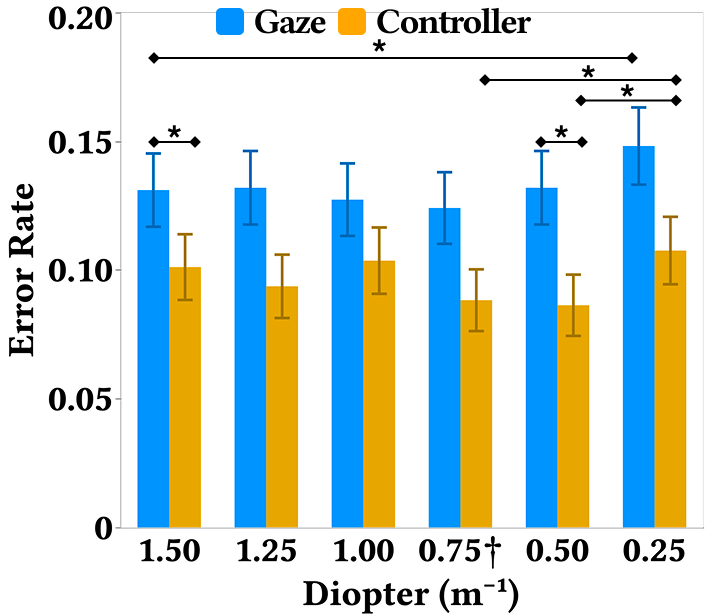}%
    }
     \hspace{0.10\textwidth}
     \subcaptionbox{\label{fig:fit}}{%
        \includegraphics[ height=0.18\textheight, keepaspectratio, alt={Fitts' law regression plot of movement time as a function of index of difficulty, with separate regression results for gaze and controller interaction.}]{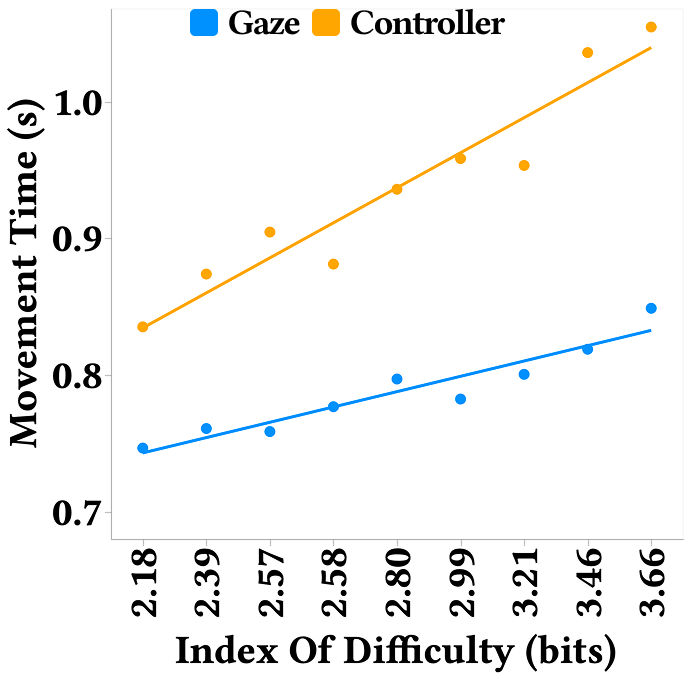}%
    }
    \caption{(a) Error rate across depth conditions (diopters) and interaction modalities. (b) Fitts’ law regression of movement time (MT) as a function of index of difficulty (ID) for \gaze and \controller.}
    \label{fig:ER_Fitts}
\end{figure*}

\subsection{Error Rate}

Error rate did not meet the normality assumption $(S = 2.41, K = 3.85)$ even after log transform; therefore, we analyzed the data using ART. The analysis revealed a significant main effect of \textbf{Depth}, $F(1, 23) = 2.462, p < .037, \eta_p^2 = .097$ (see \Cref{tab:RM-ANOVA} and \Cref{fig:ER}). Participants tended to make more errors at larger viewing distances, corresponding to lower diopter values. We also observed a significant main effect of \textbf{ID}, $F(8, 184) = 2.410, p < .017, \eta_p^2 = .095$, indicating higher error rates for more difficult tasks. Across depth conditions, gaze-based pointing produced higher error rates than controller-based pointing. In particular, error rates increased at the farthest depth condition ($0.25\,D$), suggesting greater difficulty when targets were positioned farthest from the display focal plane.

\subsection{Angular Throughput}

Angular throughput was normally distributed $(S = 0.11, K = -0.53)$. The analysis included both \gaze and \controller pointing modalities. A RM ANOVA revealed a significant main effect of \textbf{Depth}, $F(5,115) = 18.653, p < .001, \eta_p^2 = .44$, (see \Cref{tab:RM-ANOVA}). Collapsed across modalities, angular throughput decreased as targets were positioned farther from the focal plane (i.e., lower diopter values) (see \Cref{fig:THP}(a)). We also observed a main effect of \textbf{Pointing Modality}, with \gaze, $F(1,23) = 3.098, p < .05, \eta_p^2 = .221$, achieving higher angular throughput than \controller across depth levels.~Furthermore, \textbf{ID} significantly affected angular throughput, with higher ID levels yielding lower throughput. As shown separately by modality in \Cref{fig:THP}(b), \controller throughput remained relatively similar across depth conditions, whereas \gaze throughput varied more strongly, peaking near the focal distance and decreasing at greater viewing distances.

\subsection{\texorpdfstring{$\bm{SD_x}$}{SDx}}

The standard deviation along the task axis ($\mathit{SD}_x$) was normally distributed after log transform $(S = -0.23, K = -1.36)$. We observed a significant main effect of \textbf{Depth}, $F(1.536, 35.321) = 122.037, p < .001, \eta_p^2 = .841$, ( see \Cref{tab:RM-ANOVA}). Overall, pointing variability increased as targets were positioned farther from the focal plane (i.e., lower diopter values). A significant main effect of \textbf{ID} was also observed, $F(8, 184) = 28.682, p < .001, \eta_p^2 = .555$, indicating that tasks with higher difficulty resulted in greater variability along the task axis. Furthermore, a strong main effect of \textbf{Pointing Modality} was found, $F(1, 23) = 228.08, p < .001, \eta_p^2 = .990$. Across depth levels, \controller pointing exhibited significantly higher $\mathit{SD}_x$ values than \gaze pointing, indicating substantially greater spatial variability. As shown in \Cref{fig:SD}, both modalities showed increased variability at farther depths, but this increase was larger for \controller, whereas \gaze showed smaller depth-dependent changes across most levels.

\subsection{Learning Effect Analysis}

\noindent Because gaze-based 3D selection may be unfamiliar to participants~\cite{mutasim2025influence},~we examined whether performance changed over the course of the experiment.~Specifically, we analyzed movement time and error rate as a function of trial order. The analysis did not reveal a meaningful learning effect for either measure, and the depth-dependent performance trends remained consistent across the session. This suggests that the observed effects of depth on gaze-based selection were unlikely to be explained by practice or learning during the experiment.

\subsection{Fitts' Law Results}

We first evaluated how well the standard Shannon formulation of Fitts’ law ~\cite{mackenzie1992fitts} modeled all observed movement times, and found $a = 0.22$ and $b = 0.34$, with $R^2 = 0.67$, indicating a moderate fit. To examine depth-specific effects, we also fitted the model separately for each diopter level (D1–D6); the resulting coefficients and goodness-of-fit values are reported in \Cref{tab:DiopterValues}.~Across depth conditions, the slope parameter $b$ remained relatively stable, suggesting a consistent relationship between movement time and task difficulty. In contrast, the intercept $a$ varied more substantially, indicating depth-dependent changes in baseline time components not explained by task difficulty, such as perceptual processing, target localization, movement initiation, or fixation stabilization~\cite{kopper2010human, mackenzie1992fitts}. In gaze-based interaction, these baseline costs may also include oculomotor processes such as saccade planning and fixation stabilization before target selection. The observed variation in $a$ across depth conditions may therefore reflect differences in the time required to establish a stable gaze on the target. One possible contributing factor is the VAC, which has been shown to affect depth perception and oculomotor behavior in stereoscopic displays~\cite{hoffman2008vergence, kramida2015resolving}. As \gaze pointing relies on accurate and stable eye movements, this variation may reflect depth-dependent perceptual and oculomotor costs associated with the VAC~\cite{hoffman2008vergence, kramida2015resolving}.

\begin{table}
\centering
\caption{Fitts' law regression coefficients across depths.}
\label{tab:DiopterValues}
\begin{tabular}{|c|c|c|c|} \hline
Diopter & $a$ & $b$ & $R^2$ \\ \hline
${D}_1$ (1.50 D = 66.6 cm) & 0.56 & 0.10 & 0.64 \\ \hline
${D}_2$ (1.25 D = 80 cm) & 0.31 & 0.11 & 0.77 \\ \hline
${D}_3$ (1.00 D = 100 cm) & 0.82 & 0.15 & 0.44 \\ \hline
${D}_4$ (0.75 D = 133.3 cm) & 0.29 & 0.18 & 0.77 \\ \hline
${D}_5$ (0.50 D = 200 cm) & 0.44 & 0.16 & 0.73 \\ \hline
${D}_6$ (0.25 D = 400 cm) & 0.44 & 0.19 & 0.74 \\ \hline
\end{tabular}
\end{table}

When analyzed separately by interaction modality (\Cref{fig:fit}), the standard Fitts’ law model showed moderate fits for both modalities, with $R^2 = 0.63$ for \gaze and $R^2 = 0.67$ for \controller (see \Cref{tab:fitts}). The lower fit for \gaze (\Cref{fig:fit}) suggests that depth-related factors may further influence gaze-based interaction, motivating Fitts’ law extensions that account for depth variations.

First, we tried the ``Change of Target Depth'' ($\mathrm{CTD}$) model~\cite{barrera2019effect}: $MT = a + b \cdot \mathit{ID} + c \cdot \mathrm{CTD}$, where $\mathrm{CTD}$ represents the Euclidean change in target depth between successive selections. This model improved the fit for both modalities (see \Cref{tab:fitts}), increasing $R^2$ to $= 0.85$ for \gaze and $0.74$ for \controller.

Next, we evaluated the ``Variation in Diopters'' ($\mathrm{ViD}$) model proposed by Bashar et al.~\cite{bashar2025effect}: $MT = a + b \cdot \mathit{ID} + c \cdot e^{f|\mathrm{ViD}|}$. Unlike CTD, which captures spatial depth differences, ViD models changes in optical focus demand expressed in diopters and more directly reflects the visual constraints of stereoscopic displays. ViD provided the best fit across both modalities, achieving $R^2 = 0.93$ for \gaze and $R^2 = 0.84$ for \controller (see \Cref{tab:fitts}). Model comparison using the Akaike Information Criterion (AIC)~\cite{akaike1974new} and the Bayesian Information Criterion (BIC)~\cite{schwarz1978estimating} further supported ViD: it produced the lowest AIC and BIC values, with large AIC differences indicating strong support according to Burnham and Anderson~\cite{burnham2011aic}. These results suggest that incorporating optical-focus variation explains movement-time changes better than both the standard Fitts’ law model and the CTD extension, particularly for gaze-based interaction.

\begin{table*}[]
\centering
\caption{Comparison of Fitts’ law models across interaction modalities. The \textit{ViD} model achieves the best fit based on AIC and BIC values.}
\label{tab:fitts}
\resizebox{0.85\textwidth}{!}{%
\begin{tabular}{cccccc}
\hline
Model                                                                                                                                   & Interaction                & Parameters                                   & $R^2$ & AIC              & BIC              \\ \hline
\multicolumn{1}{c|}{\multirow{2}{*}{$MT = a + b \cdot \mathit{ID}$~\cite{mackenzie1992fitts}}}                             & \gaze       & $a = 0.23$, $b = 0.29$                       & 0.63  & -95.34           & -104.56           \\ \cline{2-6}
\multicolumn{1}{c|}{}                                                                                                                   & \controller & $a = 0.33$, $b = 0.12$                       & 0.67  & -102.43          & -110.45          \\ \hline
\multicolumn{1}{c|}{\multirow{2}{*}{$MT = a + b \cdot \mathit{ID} + c \cdot \mathrm{CTD}$~\cite{barrera2019effect}}}                & \gaze       & $a = -0.38$, $b = 0.29$, $c = 0.023$         & 0.85  & -194.34          & -210.10          \\ \cline{2-6}
\multicolumn{1}{c|}{}                                                                                                                   & \controller & $a = -0.27$, $b = 0.35$, $c = 0.065$         & 0.74  & -174.23          & -188.43          \\ \hline
\multicolumn{1}{c|}{\multirow{2}{*}{$MT = a + b \cdot \mathit{ID} + c \cdot e^{f|\mathrm{ViD}|}$~\cite{bashar2025effect}}} & \gaze       & $a=-0.548$, $b=0.346$, $c=0.258$, $f=-2.205$   & 0.93  & \textbf{-239.44} & \textbf{-252.542} \\ \cline{2-6}
\multicolumn{1}{c|}{}                                                                                                                   & \controller & $a=-0.38$, $b=0.42$, $c=0.715$, $f=-0.411$ & 0.84  & \textbf{-205.22} & \textbf{-215.42} \\ \hline
\end{tabular}%
}
\end{table*}

\subsection{Subjective Measure:}

\begin{figure}
    \centering
    \includegraphics[ width=\linewidth, alt={Plot summarizing NASA-TLX workload scores and System Usability Scale scores for gaze and controller interaction, enabling comparison of perceived workload and usability between the two modalities.} ]{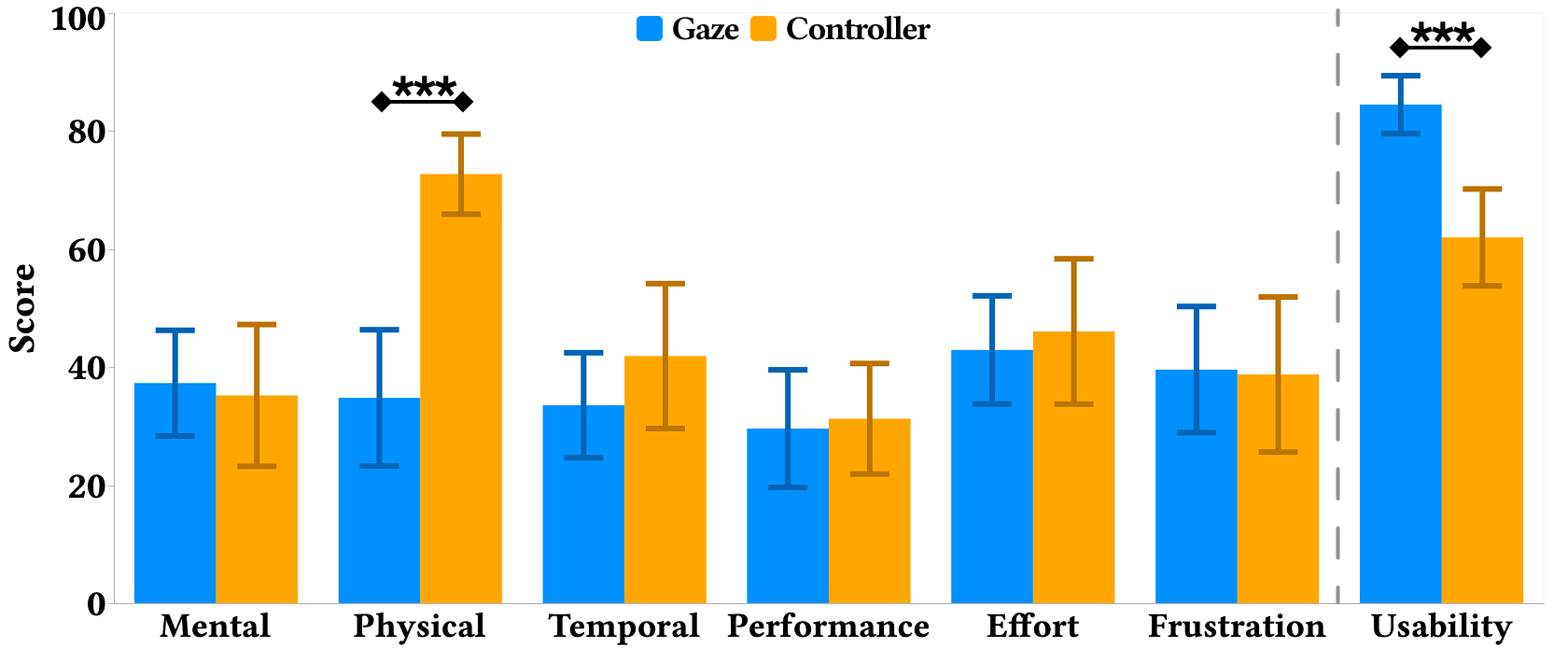}
    \caption{NASA-TLX and SUS results for \gaze and \controller interaction.}
    \label{fig:Sub}
\end{figure}

The mean SUS score was 80.31 ($\sigma = 11.92$) for \gaze and 74.69 ($\sigma = 18.61$) for \controller. A Wilcoxon signed-rank test showed that this difference was significant ($Z = -3.852$, $p < .001$), indicating that participants perceived gaze interaction as overall more usable. For workload, most NASA-TLX subscales showed no significant differences between modalities. However, physical demand was significantly higher for \controller than for \gaze ($Z = -3.851$, $p < .001$) (see \Cref{fig:Sub}), reflecting the additional motor effort required for handheld controller input. The overall NASA-TLX score was not significant between the two modalities.

\subsection{Qualitative Feedback on Depth Perception}

Participants consistently described intermediate depths as the most comfortable and easiest for selection. Depths around $1.0\,D$ (100\,cm) and $0.75\,D$ (133\,cm) were frequently perceived as balanced viewing distances. For example, P4 noted that \textit{``D4 (0.75 D) is the most comfortable because it is not too far and not too close,''} while P12 described this range as \textit{``not too far, not too close, it is just at the perfect zone.''} Other participants similarly referred to the mid-range depth as providing a \textit{``balanced depth''} (P5) or a \textit{``medium range [that] was comfortable to look at''} (P17).

In contrast, participants reported greater difficulty at farther depths, particularly D5 ($0.50\,D$, 200\,cm) and D6 ($0.25\,D$, 400\,cm), where aiming accurately or determining cursor position became harder. For instance, P2 stated that \textit{``the further away the balls were, the harder it was to see where my pointer was going,''} while P11 noted that \textit{``too far makes it difficult to aim.''}~Very close targets were also perceived as less comfortable, often requiring more movement; as P4 explained, \textit{``if it is too close, it feels like too much movement is involved to hit each target.''}~One participant additionally reported mild visual strain, noting \textit{``some pressure on my eye at the end''} (P1).~Overall, participants described intermediate depths, especially D3--D4 ($1.00$--$0.75\,D$), as the most stable and comfortable, while targets placed either too far or too close were perceived as more difficult to select.

Participants also reported low fatigue on a 7-point scale (1 = completely normal, 7 = complete fatigue). Physical fatigue was slightly higher for \controller interaction ($M = 2.38$, median = 1.5, $SD = 2.65$) than for \gaze interaction ($M = 1.04,$ median = 0, $SD = 1.68$), while mental fatigue remained low for both modalities (\controller: $M = 1.13$, $SD = 2.13$; \gaze: $M = 0.38$, $SD = 1.01$). These results suggest the task imposed minimal physical and cognitive demand, partly as participants could rest their arms between rounds.

\section{Discussion}

This study investigated how VAC-related depth variations influence \gaze-based 3D target selection in immersive virtual environments. Using an ISO~9241-411 multidirectional selection task with depth levels expressed in diopters, we evaluated \gaze pointing with \controller-based raycasting as a reference condition.

\subsection{Depth Effects on Gaze-Based Selection}

Across multiple performance measures, depth significantly affected \gaze-based target selection. Movement time increased, and angular throughput decreased as targets were placed farther from the display focal plane (i.e., at lower diopter values; see \Cref{fig:MT} and \ref{fig:THP}). Error rates also increased for distant targets, particularly at the farthest depth condition (see \Cref{fig:ER}). Together, these findings suggest that participants' \gaze pointing performance significantly decreases as targets move farther from the focal region of the display.

While participants completed a training session before the main trials, we also tested for learning effects during the study. This analysis showed no meaningful change in movement time or error rate across trial order. Thus, the observed performance differences are unlikely to be explained by practice effects and instead appear to reflect depth-related visual demands associated with the VAC.

One possible explanation is the oculomotor demands imposed by the VAC. Because gaze pointing relies directly on eye movements governed by the vergence system, changes in vergence demand across depth conditions may introduce variability during fixation shifts and target acquisition.~Vision science research has shown that the VAC can increase vergence latency and reduce convergence accuracy during stereoscopic viewing~\cite{hoffman2008vergence, vienne2020depth}.~Such oculomotor inconsistency may lead to increased corrective eye movements when attempting to stabilize gaze on targets positioned far from the display focal plane~\cite{monier2025ocular, spiegel2024vergence}.

Interestingly, performance was most consistent near the focal region of the display (approximately at $0.75\,D$). This pattern is consistent with prior work on the stereoscopic ``zone of comfort''~\cite{shibata2011zone}, which describes a range around the display's focal plane where the visual system can maintain relatively stable vergence and accommodation responses with minimal effort. When targets fall outside this region, the mismatch between vergence and accommodation increases, potentially introducing additional visual effort and reducing selection efficiency.

These observations are consistent with the participants' feedback. Participants frequently described intermediate depths (approximately $1.0\,D$ and $0.75\,D$) as the most comfortable and easiest distances. In contrast, farther-away targets were reported as more difficult to aim at, with participants indicating that it became harder to determine where the cursor was pointing. These subjective reports reinforce the quantitative findings, suggesting that gaze-based selection performance is closely tied to the perceptual comfort zone around the display focal plane.

\subsection{Comparison with Controller-Based Pointing}

Although the primary focus here is \gaze-based interaction, including \controller-based pointing as a reference condition provides useful context for interpreting the observed effects. As shown in \Cref{fig:MT} and \ref{fig:THP}, both pointing modalities were influenced by target depth, but the magnitude of these effects differed. In particular, \gaze pointing exhibited greater variations in movement time and angular throughput across depth conditions, whereas \controller-based pointing remained comparatively similar.

An additional consideration is that gaze plays a fundamental role in most pointing interactions, regardless of the input modality used for selection. In typical point-and-select tasks, users first visually fixate on a target before executing the motor action required to select it, whether with a controller, the hand, or the eyes themselves. Consequently, gaze effectively guides the planning and execution of many spatial interaction tasks. In \gaze pointing, this visual stage directly determines the cursor position, eliminating the additional motor movement required to align a controller or hand with the target~\cite{jacob1990you, majaranta2014eye}. This direct mapping between visual fixation and cursor placement may partly explain why \gaze achieved shorter movement times and higher angular throughput compared to \controller-based pointing in our study.

This performance difference also reflects the distinct motor systems involved in the two interaction techniques. Controller pointing relies primarily on hand and arm movements guided by both visual and proprioceptive feedback, which may help compensate for perceptual uncertainty in depth~\cite{batmaz2020precision, batmaz2023re, kasuga2022integration}. In contrast, \gaze pointing depends entirely on eye movements for cursor positioning. As a result, changes in vergence demand introduced by the VAC can more directly influence gaze-based target selection.

Prior research on VAC effects in manual interaction has reported increased movement time and reduced throughput for targets located at varying depths~\cite{barrera2018stereo, batmaz2022effect}. Our findings extend these observations by showing that depth-dependent performance variations are not limited to visually guided manual pointing. In \gaze-based selection, the eyes are both the perceptual system for locating the target and the input channel for controlling the cursor. As a result, vergence-related instability or increased visual demand can directly affect cursor positioning and selection performance. This distinction helps explain why depth-aware modeling, particularly the diopter-based formulation, better captured performance changes in our data. These results suggest that gaze-based XR interaction should be modeled not only as a pointing technique, but also as an interaction method constrained by oculomotor behavior under stereoscopic viewing.

Despite these depth-dependent effects, \gaze pointing achieved shorter movement times and higher angular throughput overall compared to \controller-based pointing. This result aligns with prior studies showing that gaze enables rapid target acquisition due to the speed of saccadic eye movements and the natural coupling between visual attention and target selection~\cite{luro2019comparative, sidenmark2019gaze, stellmach2012look}. However, our results also show that this advantage is modulated by target depth, highlighting the need to account for VAC-related visual constraints when designing gaze-based interaction in fixed-focus stereoscopic HMDs.

\subsection{Pointing Variability Across Depth}

The $\mathit{SD}_x$ analysis provides additional insight into how depth influences the spatial stability of gaze-based selection. As shown in \Cref{fig:SD}, pointing variability increased as targets were positioned farther from the display focal plane (i.e., at diopter values \textless 1). This suggests that VAC-related perceptual limitations may introduce additional uncertainty when users attempt to fixate and maintain a stable gaze cursor with targets located at greater depths~\cite{bashar2025effect}.

\gaze pointing directly maps eye fixation to cursor position, making cursor placement sensitive to instability in binocular fixation or vergence behavior, which can immediately influence cursor placement. Prior vision science research shows that the VAC can reduce fixation stability and induce small oscillatory eye movements when viewing objects away from the focal plane~\cite{daniel2019induced, shibata2011zone}, which likely contributes to the increased spatial variability observed in our results.

Interestingly, overall $\mathit{SD}_x$ values were lower for \gaze compared to \controller. This difference likely reflects the distinct motor systems involved: controller interaction depends on arm and hand movements, which introduce small motor deviations~\cite{mackenzie1992fitts}, whereas \gaze pointing relies on relatively small eye movements that can maintain precise fixation under stable viewing conditions~\cite{duchowski2017eye}. However, variability in \gaze pointing increased systematically with depth, indicating that gaze itself is affected by VAC-related depth conditions.

This finding highlights an important property of gaze interaction in immersive environments: although gaze enables rapid target acquisition, its spatial stability is tightly coupled to the visual system’s ability to maintain accurate binocular fixation across depths. Consequently, interface elements placed far from the focal plane may be more susceptible to gaze instability, potentially reducing selection reliability. Participant feedback supports this interpretation, as participants reported that distant targets were harder to aim at or maintain alignment with. These results also suggest that display technologies designed to reduce or eliminate the VAC, such as varifocal or multifocal displays, may improve the stability and reliability of gaze-based interaction by restoring a more natural coupling between vergence and accommodation.

\subsection{Implications for Modeling 3D Interaction}

Our modeling results further highlight the importance of explicitly incorporating depth-related visual factors when modeling gaze-based interaction performance. When applying the standard Shannon formulation of Fitts’ law~\cite{mackenzie1992fitts}, the model produced only moderate fits for both interaction modalities, with $R^2 = 0.63$ for \gaze and $R^2 = 0.67$ for \controller pointing. This result is not surprising, as classical Fitts’ law formulations were originally developed to model manual pointing tasks and therefore primarily capture motor behavior associated with hand movements~\cite{fitts1954information, soukoreff2004towards}. Depth-aware extensions of the model significantly improved the predictive accuracy, e.g., the Change of Target Depth (CTD) model~\cite{barrera2019effect} increased the model fit to $R^2 = 0.85$ for \gaze and $R^2 = 0.74$ for \controller. Similarly, the Variation in Diopters (ViD) model~\cite{bashar2025effect} produced the strongest overall fit, achieving $R^2 = 0.93$ for \gaze and $R^2 = 0.84$ for \controller. Since $R^2$ is not an appropriate metric for comparing linear and nonlinear models~\cite{spiess2010evaluation}, we relied on information criteria for model comparison. Model evaluation using AIC further confirmed these improvements ($\Delta AIC > 10$), indicating strong support for the depth-aware formulations. Specifically, the ViD model directly represents changes in optical focus demand, capturing how variations in vergence–accommodation requirements influence interaction performance.

These results suggest that movement time in gaze-based 3D selection is strongly modulated by optical depth changes rather than purely geometric distance. Because \gaze pointing directly depends on the oculomotor system, changes in vergence demand introduced by stereoscopic displays appear to play a measurable role in shaping interaction performance. Importantly, this finding goes beyond simply demonstrating that gaze is fast (\Cref{fig:MT}). Instead, it shows that gaze-based interaction exhibits systematic depth-dependent behavior that can be modeled better when depth is expressed in terms of optical demand (diopters). This insight is particularly relevant for predictive interaction modeling and adaptive interface design in XR systems. Designers and researchers developing gaze-based interfaces may therefore benefit from incorporating depth-aware models to predict selection performance or optimize interface layouts.

\subsection{Subjective Feedback and User Perception}

Subjective responses provided additional insight into participants’ experiences with \gaze interaction across depth levels. Overall, the SUS scores indicated generally positive usability, while participant comments revealed noticeable differences in interaction quality across depths. Participants reported that gaze-based pointing occasionally felt unstable during selection. Prior work has shown that natural eye-movement variability can influence gaze interaction performance~\cite{hansen2018fitts, miniotas2000application}. In our study, this instability was more noticeable when targets were positioned farther from the display focal plane, corresponding with the higher error rates (\Cref{fig:ER}) and reduced throughput (\Cref{fig:THP}) observed in the quantitative results.

Participants also reported that interaction felt more comfortable with targets at intermediate depths, whereas targets farther from the focal plane were perceived as harder to aim at or visually track.~These observations align with vision science findings that the VAC can reduce fixation stability and increase oculomotor effort when interacting with objects away from the display focal plane~\cite{akeley2004stereo, daniel2019induced, shibata2011zone}. Fatigue ratings were low overall, showing that the task was not physically or cognitively demanding. However, occasional reports of eye strain and depth-related difficulty indicate that factors related to VAC may still affect the perceived stability and comfort of gaze interaction.

Together, these observations indicate that while \gaze interaction offers advantages in speed and physical effort, its perceived stability remains susceptible to the VAC. Because most current commercial XR systems still rely on fixed-focus stereoscopic displays, these VAC-related constraints remain broadly relevant to gaze-based interaction on existing HMDs. Designers of gaze-enabled interfaces may therefore benefit from placing targets close to the focal plane of the headset, incorporating stabilization techniques, or using adaptive selection mechanisms, particularly for targets farther from the display's focal plane.

\section{Design Recommendations}

\textbf{Place frequently used gaze targets near the display's focal region.}~Gaze-based selection was most efficient near the HMDs' focal region, where movement time was lower and angular throughput higher.~Frequently used gaze-selectable elements, such as primary commands, persistent tools, or important interface panels, should therefore be placed close to the focal region when possible, within $\pm~0.5\,D$~\cite{bashar2025effect, shibata2011zone}, to support faster and more stable gaze interaction.

\textbf{Avoid precision-critical targets at large depth offsets.}~Selection performance degraded at larger depth offsets, with reduced throughput and increased error rates. Developers should therefore avoid placing small, dense, or precision-critical gaze targets far from the focal region. If such placement is unavoidable, interfaces should enlarge targets, increase spacing, or provide additional confirmation mechanisms to reduce accidental selections.

\textbf{Use depth-aware stabilization for gaze-based pointing.}~Because gaze input is directly affected by fixation stability and small involuntary eye movements, depth-dependent visual demands can reduce cursor stability. Designers should thus consider gaze-stabilization mechanisms such as temporal smoothing, adaptive filtering, target expansion, or snap-to-target assistance. These mechanisms may be particularly useful for targets outside of the focal region, where our results showed weaker gaze-selection performance.

\textbf{Consider hybrid or adaptive interaction for precision-sensitive tasks.}~Although gaze interaction enabled faster movement times and higher throughput overall, some participants perceived controller input as more precise. For tasks requiring both speed and precision, hybrid techniques may provide a better balance: gaze can be used for rapid target acquisition, while hand gestures, controllers, or other manual input can support confirmation, correction, or fine adjustment. Such adaptive approaches may be especially beneficial when targets are located at challenging depths.

\textbf{Incorporate depth-aware modeling when designing gaze interfaces.}~Our modeling results showed that depth-aware formulations of Fitts’ law, particularly the \textit{ViD} model, better explained performance variations across depth than conventional formulations. Designers and researchers should thus account for optical depth, not only geometric distance, when predicting gaze-selection performance or designing adaptive XR interfaces. For example, depth-aware models could inform dynamic target sizing, placement optimization, or interaction-technique switching across depth.

\section{Limitations \& Future Work}

This study was conducted using a single fixed-focus HMD, reflecting interaction under conventional stereoscopic viewing conditions. Although fixed-focus stereoscopic displays remain common in current commercial XR systems, the magnitude and spatial pattern of the observed depth-dependent effects may still differ across devices depending on focal distance, optical design, eye-tracking precision, field of view, and calibration stability. Future work should therefore replicate this study across HMDs with different focal distances, optical configurations, and eye-tracking systems. The findings may also not generalize to emerging display technologies designed to reduce or eliminate the VAC, such as varifocal~\cite{hu2026varifocaldisplaysreduceimpact}, multifocal~\cite{batmaz2022effect}, or light-field displays~\cite{maeda2024wide}. Future work should examine whether similar depth-dependent behavior persists when the VAC is reduced or eliminated.

The experiment also employed a controlled ISO~9241-411 multidirectional selection task, which enabled comparison with prior pointing studies~\cite{barrera2019effect, bashar2025effect} but does not capture the full range of gaze-based interactions used in XR. Although selection is a fundamental component of many 3D user interface interactions~\cite{laviola20173d}, future work should examine whether the observed depth-dependent effects generalize to more complex tasks such as navigation and object manipulation~\cite{medeiros2023benefits}.

Finally, the explored depth range corresponds to typical distances used for floating interface elements in current XR systems. Although the variation in depth was moderate, measurable performance differences were observed within this range. Future studies should examine wider depth separations, multi-plane interfaces, and longer-term use to better understand how gaze interaction adapts to VAC-related depth constraints over time.

\section{Conclusion}

This work investigated how the vergence-accommodation conflict (VAC) influences gaze-based 3D target selection in immersive virtual environments. Our results show that depth significantly affects gaze-based interaction performance: as targets move farther from the display focal plane, movement time and error rates increase, while throughput decreases. These findings indicate that gaze-based pointing is prone to depth variations introduced by the VAC. Despite its susceptibility to noise, gaze interaction achieved faster movement times and higher overall throughput than controller-based pointing, highlighting its potential as a rapid pointing modality in immersive systems. Furthermore, modeling results showed that incorporating variation in diopters (ViD) substantially improves the predictive power of Fitts’ law models. By focusing on gaze-based selection, this work extends prior VAC research from manual pointing to an interaction modality in which the oculomotor system itself serves as the pointing channel. Together, these findings highlight the need for depth-aware design and modeling of gaze-driven XR interfaces.

\acknowledgments{The illustrated participant depictions in the teaser figure were refined using OpenAI’s ChatGPT (GPT-5.4) and subsequently reviewed and edited by the authors. All other elements of the figure remain the original work of the authors.}

\bibliographystyle{abbrv-doi-hyperref}

\bibliography{template}

\end{document}